\documentclass[12pt,preprint]{aastex}
\usepackage{epsfig}
\shorttitle{IR Continuum of Be Star Disks}
\shortauthors{Touhami et al.}

\begin{document}

\accepted{}

\title{The Infrared Continuum Sizes of Be Star Disks}

\author{Y. Touhami, D. R. Gies}
\affil{Center for High Angular Resolution Astronomy and
 Department of Physics and Astronomy,
 Georgia State University, P. O. Box 4106, Atlanta, GA 30302-4106;
 yamina@chara.gsu.edu, gies@chara.gsu.edu}

\author{G. H. Schaefer}
\affil{Georgia State University, CHARA Array, P.\ O.\ Box 48,
Mount Wilson, CA 91023; schaefer@chara-array.org}



\begin{abstract}
We present an analysis of the near-infrared continuum emission
from the circumstellar gas disks of Be stars using a radiative
transfer code for a parametrized version of the viscous decretion disk model.
This isothermal gas model creates predicted images that
we use to estimate the HWHM emission radius along the major axis
of the projected disk and the spatially integrated flux excess
at wavelengths of  1.7, 2.1, 4.8, 9, and 18~$\mu$m.
We discuss in detail the effect of the disk base density,
inclination angle, stellar effective temperature, and other
physical parameters on the derived disk sizes and color excesses.
We calculate color excess estimates relative to the stellar
$V$-band flux for a sample of 130 Be stars using photometry
from 2MASS and the AKARI infrared camera all-sky survey.
The color excess relations from our models make a good
match of the observed color excesses of Be stars.
We also present our results on the projected size of the
disk as a function of wavelength for the classical
Be star $\zeta$~Tauri, and we show that the model predictions
are consistent with interferometric observations in the
$H$, $K^\prime$, and 12~$\mu$m bands.
\end{abstract}



\section{Introduction}                              

Be stars are rapidly rotating B-type stars that at times host a
circumstellar gas disk \citep{por03}.  They exhibit various
observational attributes such as photospheric non-radial pulsations,
hydrogen and iron emission lines that are formed in the disk,
infrared, millimeter, and radio continuum excess emission from the disk,
intrinsic linear polarization from electron scattering in the disk, and,
in an increasing number of cases, the presence of a compact
companion \citep{str31,sle88,por03}.
The circumstellar ionized gas around Be stars gives rise to a large
infrared excess emission that is seen in their spectral energy distribution.
Observations conducted by \citet{woo70} at 10~$\mu$m
were the first to suggest that the IR excess of Be stars is
due to free-free and bound-free emission and not due to
thermal dust emission.  An early study by \citet{geh74} confirmed
that the large IR emission detected from 33 Be stars originates in the
free-free emission only and that thermal dust emission fails
to reproduce the observations at 20~$\mu$m.  The IR excess from
ionized disk gas was subsequently investigated with space-based data from
IRAS \citep{cot87,wat87,dou94}, ISO \citep{wat00}, and MSX \citep{cla05},
and with ground-based mid-IR \citep{rin99} and
near-IR data from the 2MASS survey \citep{zha05}.
The near- and mid-IR excess flux emission is known to increase
with wavelength and to dominate over the stellar flux at long wavelengths.
Observations probing the circumstellar disk structure
at many wavelengths and across all Be spectral types are necessary to
understand the complex physical processes involving disk formation.

Models of the disk continuum emission have become increasingly
comprehensive in recent years.  Some of the first models that assumed
a simple geometry and gas density distribution \citep{wat86,kas89,dou94,rin99}
were successful in fitting the color excesses in many cases.
A significant improvement came with the development of the viscous
decretion disk model \citep{lee91,por99,oka01} in which the gas
orbits with nearly-Keplerian velocity, but with a small radial
outflow motion caused by turbulent viscosity.  These models characterize
the disk density with a radial power law, $\rho \propto R^{-3.5}$,
and a vertical extension set by hydrostatic equilibrium.
The IR-excesses predicted are similar to those observed, but often
require a density exponent with smaller absolute value and/or
a non-isothermal temperature distribution in order to
match observations \citep{por99,prt03}.
Recent models include radiation processes in the disk and
determine fully consistent temperature and density distributions
\citep{cab06,jon08}, and these are remarkably successful in explaining
the continuum and line emission in some Be stars \citep{car06,hal08}.

The advent of optical long baseline interferometry has now led
to the direct resolution of Be star disks in both emission lines
\citep{qui97,ste01,tyc06,mei07,car09} and in the IR-continuum \citep{gie07,mei07,mei09}.
These observations offer us the means to explore the geometry of Be disks and to
measure the disk gas temperature.  Since the IR-excess flux
depends mainly on the projected size of the disk and the gas surface
brightness (from the temperature dependent source function),
a comparison of the angular size and flux excess provides
a way to investigate the disk-to-star temperature ratio and
(in principle) to study spatial temperature variations in the disk.

Our goal in this paper is to explore the IR-excess flux predictions
for a parametrized version of the isothermal, viscous disk
model that we have developed to study CHARA Array $K^\prime$-band observations
of Be disks \citep{gie07}.  We calculate the near-IR color excesses
for various representative stellar and disk parameters and
compare these with new IR measurements from the AKARI satellite \citep{ita10}.
In \S2, we describe the model in detail and discuss how the different
physical parameters of the model influence the color excess. In \S3,
we discuss the color-color diagrams and compare our results to the AKARI
all-sky survey observations of Be stars. Finally in \S4, we consider the
example of the Be star $\zeta$~Tau, and we show that the predicted
variation of circumstellar disk angular size with wavelength is consistent
with the available multi-wavelength observations.


\section{Be Star Circumstellar Disk Models}         

Our original model \citep{gie07} created images of Be stars and
their disks for comparison with $K^\prime$-band interferometry
from the GSU CHARA Array.  In order to investigate the disk emission over a
range of wavelengths in the near and mid-IR, we have extended the model to calculate the disk flux
distribution at 1.66 ($H$-band), 2.13 ($K^\prime$-band), 4.8, 9, and 18 $\mu$m.
Our code is a realization of an isothermal viscous decretion disk model.
In brief, the gas density distribution is given by
\begin{equation}
\rho (R,Z) = \rho_0 R^{-n} \exp \left[-{1\over2}\left({Z\over{H(R)}}\right)^2\right]
\end{equation}
where $R$ and $Z$ are the radial and the vertical cylindrical coordinates,
respectively, in units of stellar radii,
$\rho_0$ is the gas base density, and $n$ is the radial density exponent.
$H(R)$ is the disk vertical scale height defined by
\begin{equation}
H(R) = \frac{c_s}{V_K} R^{3/2}
\end{equation}
where $c_s$ is the sound speed (in turn dependent on the disk gas
temperature) and $V_K$ is the Keplerian velocity at the stellar equator.
Our multi-wavelength disk model has four physical parameters: the base density
$\rho_0$, the density exponent $n$, the disk-to-star temperature ratio $T_d / T_{\rm eff}$,
and the outer boundary disk radius $R_{\rm out}$. There are also two observational parameters:
the wavelength $\lambda$ and the inclination angle $i$.


The code follows the method of \citet{hum00} and solves the equation of
transfer along a grid of sight-lines around the star and disk.  The result is
a spatial image of the system (see examples in \citealt{gie07}).
We assume source functions equal to the Planck functions
for the temperatures of the star and disk.
The disk optical depth in the near-IR is mainly due to free-free and
bound-free processes, and it can be expressed using an incremental
step $ds$ along a projected rectilinear coordinate grid as follows
\begin{equation}
d\tau = C(\lambda, T_d) \rho (R,Z)^2 ~ds
\end{equation}
where the coefficient $C(\lambda, T_d)$ is given by equation (5) in \citet{dou94}.
This coefficient includes terms for the Gaunt factors for bound-free and free-free emission.
We set these Gaunt factors by interpolating in wavelength in the tables from \citet{wat84}.
For simplicity, we estimate the ionization equilibrium and Gaunt factors
throughout the disk for two idealized cases
(adopted from \citealt{lam84} and \citealt{wat84}): (1) a hot plasma with
ionized H, singly ionized He, and doubly ionized C, N, O for disk temperatures above 15000~K, and
(2) a warm plasma consisting of ionized H, neutral He, and singly ionized C, N, and O for
lower temperatures.  Note that we ignore the electron scattering flux in the model,
since this flux source is proportional to the stellar flux and hence is relatively small
at longer wavelengths.

Here we focus on two results of the calculation, the flux excess and
apparent disk size.  We determine from the derived image the total monochromatic flux of the
star and disk, $F_{\rm total}$, and then we estimate the net disk contribution
as $F_d=F_{\rm total}-F_s$, i.e., the net flux relative to the unobscured star.
We show this quantity in a magnitude form of flux excess as
$E^\star (V^\star - m_\lambda) = 2.5\log (1+F_d/F_s)$.
This notation is based upon the assumption that the disk contribution is
negligible in the optical $V$-band, so that we can write the flux
excess as how many magnitudes brighter the system appears at longer
wavelengths where the disk shines brightly.
This latter assumption should be taken
with caution. In fact, there is ample evidence
from the $V$-band brightening of Be stars during their active
phases that the disks do contribute to the $V$-band
(perhaps by as much as $50\%$) in those cases with dense and
large circumstellar disks. Consequently, by referring the
colors to a $V$ magnitude that is brighter than that for the star
alone, the observed color excesses may be lower than the calculated
color excesses in some cases.  The star superscript is used here to
differentiate this type of ``reddening'' or color-excess from the kind
normally associated with interstellar extinction.   \citet{wat86} expresses
the monochromatic flux excess as $Z = F_{\rm total} / F_s$, so in
our notation $E^\star (V^\star - m_\lambda) = 2.5\log Z$.
We also use the calculated
spatial image of the star plus disk to find the HWHM of the emission envelope
along the projected major axis of the disk, and we define an effective,
observational disk radius $R_d / R_s$ as the ratio of the angular HWHM
to the angular stellar radius.

We begin by showing our results on the flux excess and disk size for a
default model, and then we show how changes in the physical and observational
parameters affect the results.  In the default model,
we assume that the central star is an early-type star with effective temperature
$T_{\rm eff} =30$~kK, radius $R_s /R_\odot = 10$, and mass $M_s /M_\odot = 15.5$.
The infrared excesses derived from observations suggest that the power-law density
exponent of the circumstellar disk falls in the range $n \approx 2.0$ to $3.5$
\citep{cot87,wat87}, so we assumed $n = 3$ here, a value consistent with
our prior interferometric results \citep{gie07}.  The other adopted parameters
for the default model are a disk-to-star temperature ratio $T_d / T_{\rm eff} = 2/3$,
an outer boundary disk radius $R_{\rm out}/R_s = 21.4$, and an inclination angle
$i=45^\circ$.  These parameters are selected from an earlier model for
the Be star $\gamma$~Cas \citep{gie07}, and the outer boundary, for example,
corresponds to the Roche radius of this binary system.
Our results for the color excess and disk radius are listed
in Table~1 as a function of waveband and base density for this model and
several others described below.

\placetable{tab1}         

We show in Figure~1 how the color excesses $E^\star(V^\star-K)$,
$E^\star(V^\star-9\mu{\rm m})$, and $E^\star(V^\star - 18\mu{\rm m})$
vary as a function of the disk base density $\rho_0$.
The disk flux excess is highly dependent on density $\rho_0$ and wavelength $\lambda$.
The color excesses at low densities are insignificant because the disk is
optically thin in the continuum and the stellar photospheric flux dominates.
As the density increases, the color excess resulting from the disk emission
becomes more important and dominates at longer wavelengths. In fact, the excess emission at
18~$\mu$m is higher than at 9~$\mu$m and at 2.13~$\mu$m because the optically thick-thin boundary
of the disk becomes bigger and the excess flux larger at longer wavelengths.

\begin{figure}
\begin{center}
{\includegraphics[angle=90,height=12cm]{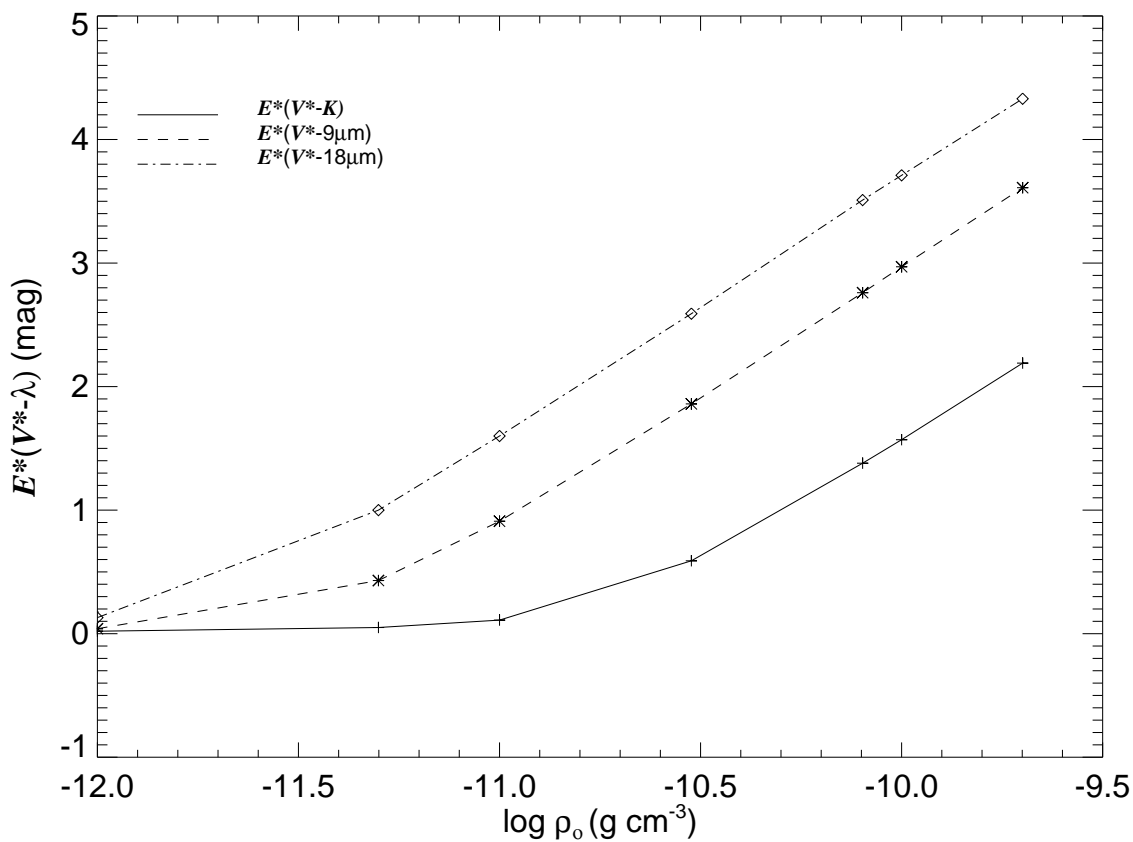}}
\end{center}
\caption{Plots of the variation in color excess as function
of disk base density for the default model parameters
($n = 3$, $i = 45^\circ$, $T_{\rm eff} = 30$ kK, $R_{\rm out} = 21.4 R_s$, $T_d = 2/3 T_{\rm eff}$).
The excess emission increases with increasing disk base density $\rho_0$
and wavelength.}
\label{fig1}
\end{figure}

\placefigure{fig1}       

The relationship between the 18~$\mu$m color excess and apparent disk emission radius
for the default model is shown as a solid line in Figure~2 (for $T_d / T_{\rm eff} = 2/3$).
We also show a number of other models where different parameters are varied in turn.
The case with a higher inclination angle $i = 80^\circ$ is plotted with a dashed line.
As the inclination increases, the projected disk surface area decreases and hence the
disk flux excess also declines.  However, at higher inclination, a ray through the
outer part of the disk encounters significant density over a longer path, and the
increased optical depth causes the effective radius to appear larger.  Consequently,
as the inclination increases, points on the default model curve are shifted to lower
color excess and higher effective radius.

\begin{figure}
\begin{center}
{\includegraphics[angle=90,height=12cm]{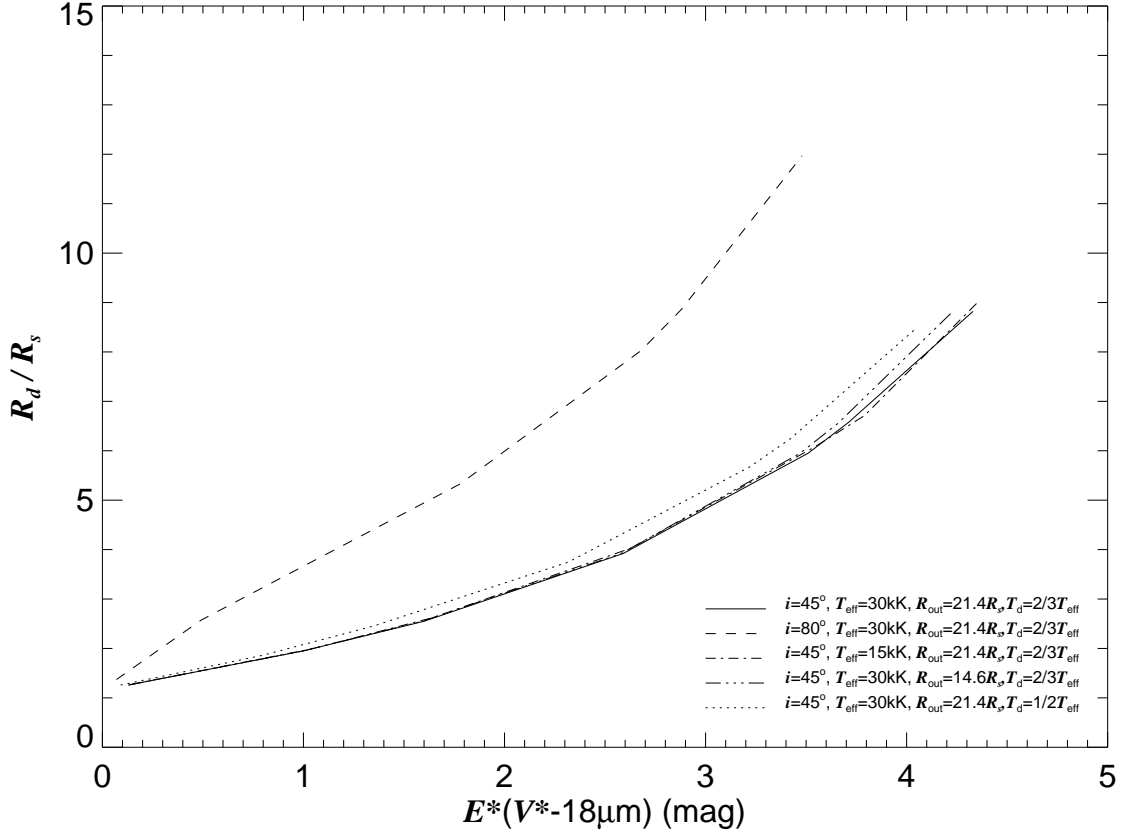}}
\end{center}
\caption{Plots of the ratio of disk HWHM radius to stellar radius as a function of the
flux excess at 18~$\mu$m.  The solid line illustrates the relationship for the
default model while the other line styles show the results found by changing
one of the model parameters (indicated in the legend).}
\label{fig2}
\end{figure}

\placefigure{fig2}  

Changes in the other parameters have less influence on the size -- color excess
relation shown in Figure~2.  For example, for the case of a cooler Be star with an
effective temperature of $T_{\rm eff} = 15$~kK (shown by the dashed-dotted line),
the resulting color excess and disk radius are almost identical to that for the default
case ($T_{\rm eff} = 30$~kK; solid line) at all values of the disk base density.
This is due to the temperature dependence of the optical depth coefficient,
$C(\lambda, T_d)\propto T_d^{-1/2}$.  The radius of the optically thick-thin boundary
will vary with this optical depth term, and the source function is
proportional to temperature in the Rayleigh-Jeans part of the spectrum. Thus,
the disk emission flux will vary as the product of projected area and source function,
or $\sim R_d^2 ~S \sim (T_d^{-1/2})^2 ~T_d$, which is approximately constant, all other
parameters being equal.

If we adopt a smaller value of the outer radius of the disk
($R_{\rm out} / R_s = 14.6$; shown by the dash-triple dotted line),
we see that there is no difference between this case
and the default case ($R_{\rm out} / R_s = 21.6$) at low densities.
It is only at very high density that the truncation of the outer disk leads
to a slight decline in the color excess.

Finally, the dotted line shows the influence of our choice of disk temperature.
In the default model, we adopt an isothermal disk with a temperature of
$T_d= 2/3 ~T_{\rm eff}$ following the example of \citet{hum00}.
We compare in Figure~2 the color excess for two disk temperatures,
$1/2 ~T_{\rm eff}$ and $2/3 ~T_{\rm eff}$.
The source function varies approximately linearly with disk temperature $T_d$,
so a drop of 25\% in disk temperature will create a comparable decrease in
emission flux (for a given disk radius).  This decrease of $\approx 0.3$ mag in
$E^\star(V^\star - 18\mu{\rm m})$ is seen in Figure~2 at the high density end
where the disk flux dominates.  This suggest that
coordinated interferometric and IR excess observations are potentially
an important means to study the disk temperature properties, especially
for Be stars with dense and large disks.

Our results show many similarities to the color excesses derived by
\citet{dou94} from a much simpler model.  They confined the emitting gas
to a wedge-like disk with a density law dependent only on the distance to the star,
and they solved the radiative transfer problem for just $i=0$ and $i=90^\circ$.
\citet{dou94} also investigated the dependence of the near-IR color
excess on the model parameters, but they present relative color excesses between
adjacent near-IR bands ($CE(J-K)$ and $CE(K-L)$; see their Fig.~7) rather than
referencing the excess to the stellar $V$-band flux as we do here.
Nevertheless, the agreement is reasonable, and, for example, their expression for the
disk optically thick-thin boundary as a power law function of wavelength
and base density (their eq.~7) is similar to our results for $R_d / R_s$ in
Table~1.  Furthermore, \citet{dou94} also presented instructive results for a range in
the density power law exponent, with values greater and less than the $n=3$ value
assumed here (they used the symbol $\beta$ for the density exponent).  Their work showed
that a smaller $n$ yields a more spatially extended disk and consequently larger color excesses.

Recent models for Be star disks present detailed calculations of the
gas temperature as a function of disk position.
\citet{cab06} solve for the temperature distribution in the disk through
a Monte-Carlo treatment of radiative transfer, and they find that gas
temperature reaches $T_d \approx 0.6 ~T_{\rm eff}$ in optically thin
parts of the disk in one representative model.
\citet{sig09} present results for a grid of models that maintain
hydrostatic equilibrium with the spatial variations in temperature,
and they find that the density averaged temperature is
$T_d \approx ~0.6 T_{\rm eff}$, however, this average temperature declines
with increasing disk density (perhaps by $20-30\%$; see their Fig.~8).
Their models indicate that the temperatures tend to be cooler in
the denser regions closer to the star and in the mid-plane.
The IR-excesses calculated from such models are similar to those
for isothermal models.  For example, \citet{cab06} show that
the predicted flux excesses for different inclination angles
are almost the same for the isothermal and non-isothermal model
they study (see the upper panel of their Fig.~10).
However, since the disk gas temperatures decline with increasing density,
we suspect that a plot like Figure~2 of color excess and radius for such
fully consistent, non-isothermal models would look similar to our
$T_d= 2/3 T_{\rm eff}$ curve at small excess (low density), but would tend
towards the $T_d= 1/2 T_{\rm eff}$ curve at large excess (high density).


\section{AKARI IR Fluxes of Be Stars}               

With the recent release of the AKARI/IRC mid-infrared all-sky survey
\citep{ish10}, we now have the opportunity to compare the observed
and model flux excesses at 9 and 18~$\mu$m.  We took as our
sample 130 Be stars from the work of \citet{dou94} that have reliable
estimates of interstellar reddening $E(B-V)$.  We then collected
$V$, $K_s$, $m(9~\mu{\rm m})$, and $m(18~\mu{\rm m})$ magnitudes from
\citet{ita10} for each target.  The $V$ magnitudes were taken
from the SIMBAD database, $K_s$ magnitudes from 2MASS \citep{skr06},
and $m(9\mu{\rm m})$ and $m(18\mu{\rm m})$ magnitudes from AKARI.
These last two are based on the Vega magnitude scale, where a
model spectrum by R.\ L.\ Kurucz defines the flux zero point
as a function of wavelength \citep{tan08}.
The photometry we use here was collected at different times,
and since Be stars are inherently variable \citep{por03},
some scatter must be expected in the results because
the flux excesses will change with disk density variations.
We also collected
stellar effective temperatures for each target from the work of
\citet{fre05} in order to estimate the intrinsic stellar colors.

Our goal is to determine a flux excess by comparing
the observed and intrinsic stellar colors of the targets.
Using the same magnitude notation given in the previous section,
we determine the near-IR flux excesses by
$$E^\star(V^\star-m_\lambda) =
  V - m_\lambda - E(B-V) \times (3.10 - R_\lambda)
  - (V - m_\lambda)({\rm Kurucz})$$
where the ratio of interstellar extinction to reddening is
$R_\lambda = A_\lambda / E(B-V)$ \citep{fit99} and
$(V - m_\lambda)({\rm Kurucz})$ is the intrinsic stellar
color derived from monochromatic sampling of flux ratios
of model spectra with the Vega spectrum from R.\ L.\ Kurucz.
These model spectra are also from Kurucz atmospheres for
solar metallicity, gravity $\log g = 4.0$, and a microturbulent
velocity of 2 km~s$^{-1}$ (parameters appropriate for main
sequence B-stars).  The intrinsic colors are listed in
Table~2 as a function of effective temperature $T_{\rm eff}$,
and we found that they made a reliable match to the AKARI
colors of B-stars with known $T_{\rm eff}$ from interferometry
and bolometric luminosity \citep{cod76}.  The derived
color excesses for the Be stars are listed in Table~3.  The typical
errors in the AKARI magnitudes are $\pm 0.05$~mag, but they can be
larger for the 2MASS $K_s$ magnitudes since many of the Be stars
are bright and their magnitudes were determined from the
the wings of the point spread function.
We caution that the intrinsic colors from Table 2 may not be
appropriate in some cases because rotational gravity darkening
will make stars with large inclination appear redder (although
the color difference may be negated if opaque disk gas blocks
the cooler equatorial zones from view).  We have ignored these
complications because they are difficult to estimate accurately and
because these colors are not too sensitive to temperature
for hot stars.

\placetable{tab2}      

\placetable{tab3}      

We plot the results in two color-color diagrams in Figures 3 and 4.
These show the flux excesses at 9 and 18~$\mu$m, respectively,
as a function of the $K_s$-band excess.  Also shown in these
figures are plots of our model near-IR excesses (\S2) for the
default model and cases with differing stellar temperature
and disk inclination.  All these models make similar predictions
about color-color excesses, and they appear to match the
observations well, especially if allowance is made for the
slight negative shift in observed colors caused by our neglect of
disk flux in the $V$-band.  There are three objects plotted
in Figures 3 and 4 with $E^\star(V^\star-K) \approx 1.0$ that
fall well below the predicted trends.  All three are
binary stars in which the companion is a K-giant
(HD~45910 = AX~Mon, \citealt{eli97}; HD~50123 = HZ~CMa, \citealt{ste94};
HD~50820, \citealt{gin02}), and we suspect that their
relatively large brightness in $K_s$ is due to the flux from
the cool giant companion.  Otherwise, the overall agreement suggests
that the viscous decretion disk model provides a satisfactory
description of the near-IR flux excesses.

\begin{figure}
\begin{center}
{\includegraphics[angle=90,height=12cm]{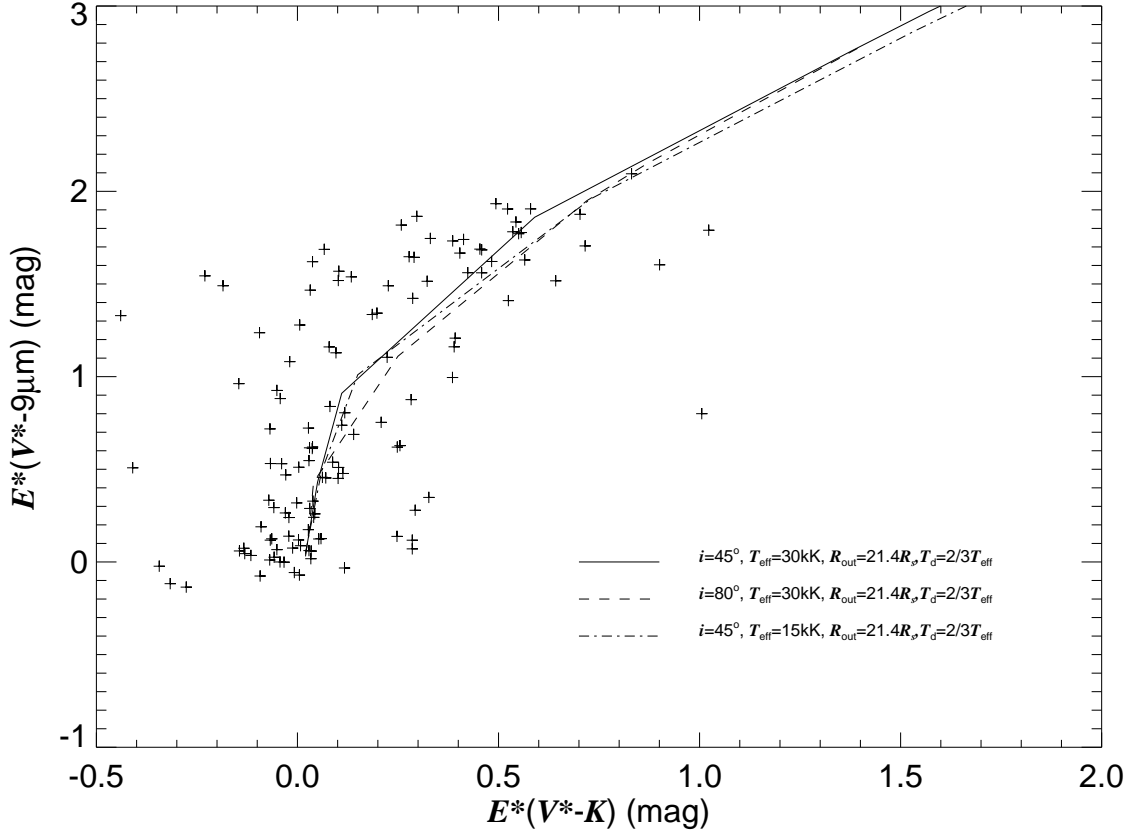}}
\end{center}
\caption{A color excess -- color excess diagram showing the 9~$\mu$m flux excess
as a function of the $K_s$-band excess. The solid line shows our results using
the default model input parameters, the dashed line represent a model with
a high-inclination disk ($i = 80^\circ$), and the dashed-dotted line shows the case of a
cooler Be star ($T_{\rm eff} = 15$ kK).  The plus signs indicate the observed excesses
of Be stars derived from 2MASS and AKARI/IRC photometry.}
\label{fig3}
\end{figure}

\placefigure{fig3}  

\begin{figure}
\begin{center}
{\includegraphics[angle=90,height=12cm]{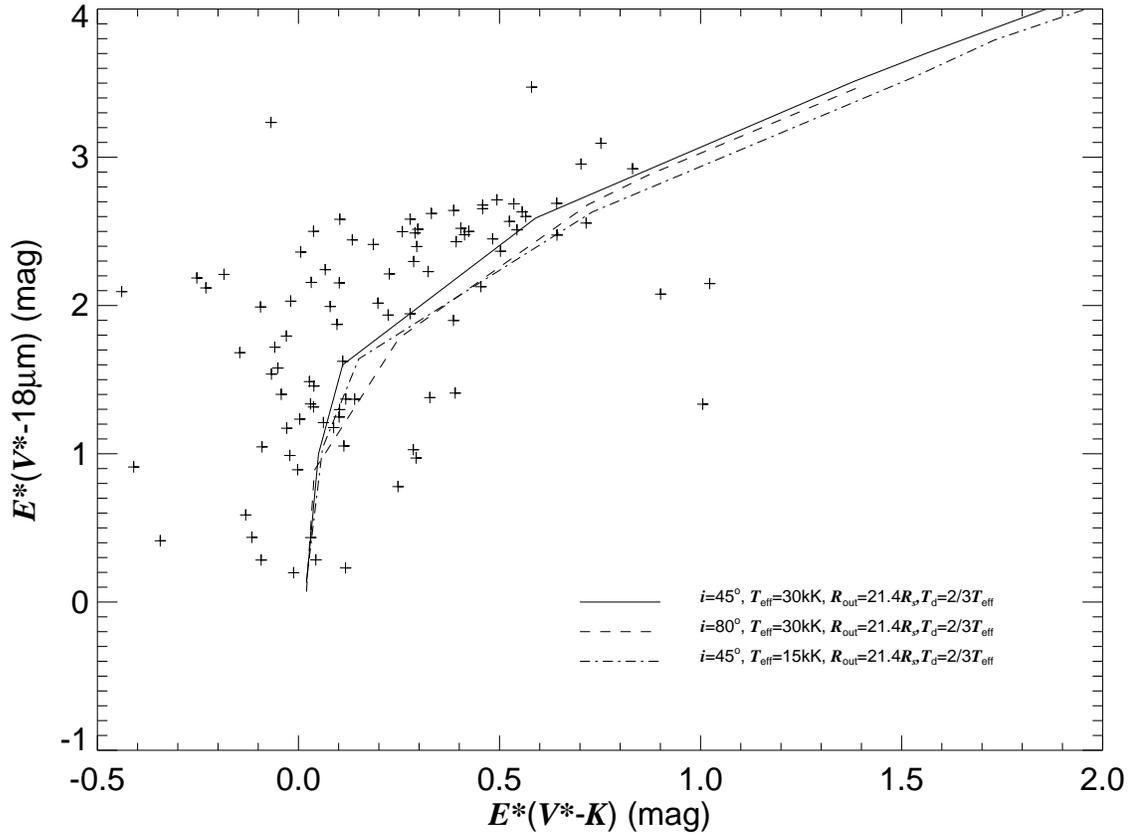}}
\end{center}
\caption{A color excess -- color excess diagram showing the 18~$\mu$m flux excess
as a function of the $K_s$-band excess (in the same format as Fig.~3).}
\label{fig4}
\end{figure}

\placefigure{fig4}  

Inspection of Figures 3 and 4 shows that at the high density
limit, the color-color excess diagrams assume a linear form.
This part of the relation occurs when the disk flux dominates
over the stellar flux and the colors become those of the disk.
Thus, we expect that all Be stars will appear with approximately the
same near-IR color for those cases with sufficiently dense disks.
As an example, we show in Figure~5 a near-IR color-magnitude
diagram for all those sample stars with Hipparcos parallax data
yielding absolute magnitude errors less than 0.5 mag \citep{van07}.
This plot shows the interstellar extinction corrected, absolute
$K_s$ magnitude versus an interstellar reddening corrected,
color index $K_s - m(9~\mu{\rm m})$ (shown as plus signs).
Each of these is connected by a dotted line to a color and
magnitude coordinate corresponding to one with the derived
color excess removed (nominally for the star itself; shown as
diamonds).  Also plotted is the zero-age main-sequence
(left dashed line) from the work of \citet{lej01} that was formed
from the theoretical $(T_{\rm eff}, V)$ track and the colors
in Table~2 for $T_{\rm eff} = 10 - 30$~kK.  From Figure~3,
the asymptotic form of the excess relation is
$$E^\star(V^\star-9~\mu{\rm m}) \approx E^\star(V^\star- K_s) + 1.35$$
and we also plot the main-sequence translated in color
by this expression and brighter in $K_s$ by 1 magnitude (see Fig.~1)
to represent the approximate positions of the dense disk case (right dashed line).
We see that the Be stars appear over a range in color between
the no disk and strong disk cases, with many of the Be stars
having colors close to the strong disk limit.  This diagram is
similar in appearance to that for the Be stars discovered in
the LMC by \citet{bon10}, who present a plot in the $(J-[3.6\mu{\rm m}], J)$
plane.

\begin{figure}
\begin{center}
{\includegraphics[angle=90,height=12cm]{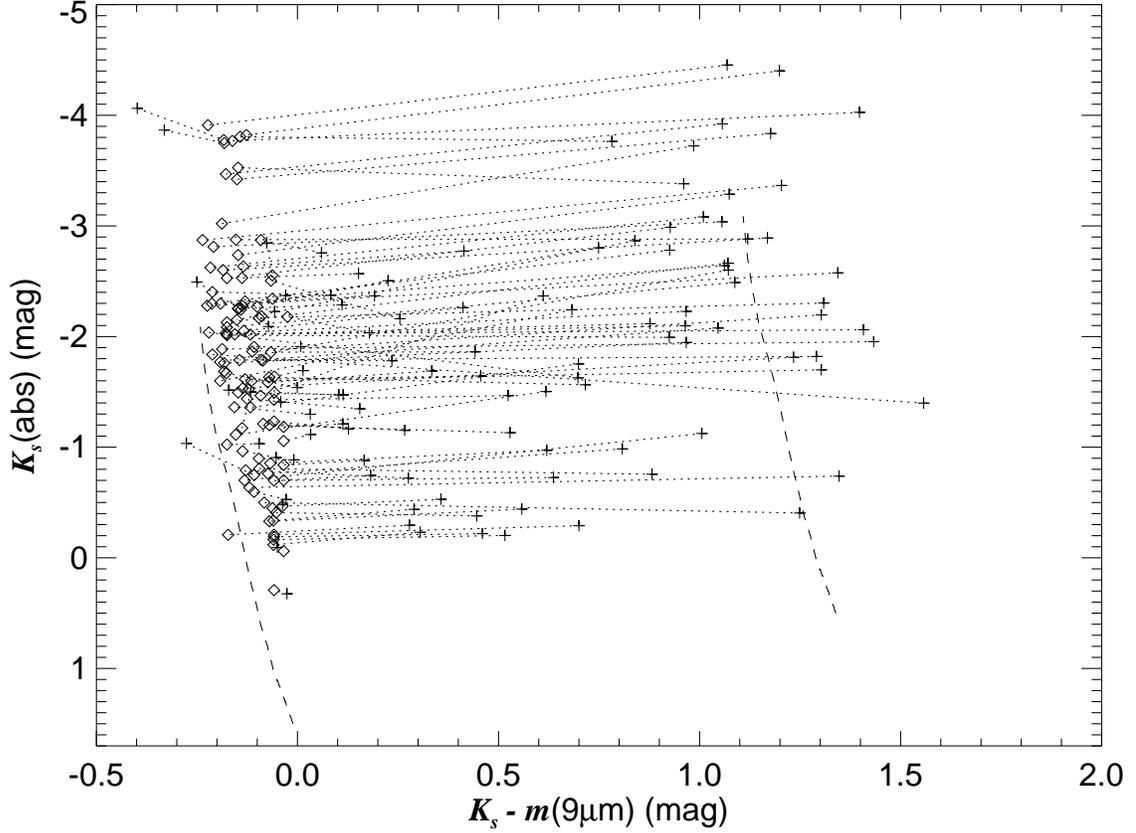}}
\end{center}
\caption{A color -- magnitude diagram showing the absolute $K_s$ magnitude
corrected for interstellar extinction as a function of the interstellar
reddening corrected, color index $K_s - m(9~\mu {\rm m})$ for our Be star
sample (shown as plus signs).  Each target point is connected by
a dotted line to a diamond representing the color and magnitude
of the B star alone.  The left dashed line represents the zero-age
main-sequence for stars with $T_{\rm eff} = 10 - 30$~kK, and
this also is shown translated in color and magnitude for a dense
disk as the dashed line on the right hand side.}
\label{fig5}
\end{figure}

\placefigure{fig5}  


\section{Angular Size of the Disk of $\zeta$ Tau}   

\citet{mei09} measured some of the first Be disk diameters at
8 and 12~$\mu$m using the VLTI and MIDI instrument.  Their
results suggested that the disk sizes do not increase with
wavelength as predicted by simple models.  For example,
they found that the angular size of the disk of the Be star
$\alpha$~Ara was approximately constant between 2 and 12~$\mu$m,
and they speculated that the disks may be truncated by the tidal
effects of a binary companion.  Our models and those of \citet{dou94}
suggest that the color excesses and effective radii could reach
finite limits in those cases with high disk density and a
small outer boundary.

Here we demonstrate that at least for one case, $\zeta$~Tau, the
angular size variation with wavelength is consistent with model predictions.
The Be star $\zeta$~Tau (HD 37202, HR 1910, HIP 26451) is a
frequently observed target with a strong IR-excess \citep{tou10}.
The H$\alpha$ emission line in its spectrum shows cyclic $V/R$ variations on
a timescale of few years, which are explained by the presence of a
one-armed oscillation in its circumstellar disk \citep{oka02,car09}.
The star is the primary in a spectroscopic binary with an orbital period of
$P = 133$~d \citep{ruz09}. The system is composed of a $11 M_{\odot}$ primary
Be type star and a $1.3 M_{\odot}$ secondary star \citep{flo89}.
Several interferometric studies of the H$\alpha$ emission line \citep{qui97,tyc04}
and the IR-continuum \citep{gie07,car09,mei09} have resolved the circumstellar disk
around the primary Be star.  A recent CHARA Array investigation by \citet{sch10}
shows that the disk of $\zeta$~Tau is viewed almost edge-on and that the
disk may be precessing with the $V/R$ cycle.

We used our model to estimate $R_d / R_s$ for $\zeta$~Tau over the
wavelength range of 1.7 -- 18~$\mu$m.  We adopted the stellar parameters from
\citet{gie07}: mass $M_s = 11.2 M_{\odot}$, radius $R_s = 5.5 R_{\odot}$,
effective temperature $T_{\rm eff} = 19$~kK, and parallax $\pi = 7.82$~mas.
The disk parameters assumed are $\rho _0 = 1.4 \times 10^{-10}$ g~cm$^{-3}$,
$n = 2.9$, $i = 87^\circ$, and $T_d / T_{\rm eff} = 2/3$ (based on fits
of recent $K^\prime$ observations with the CHARA Array).
The outer disk boundary was set at the binary Roche radius,
$R_{\rm out} = R_{\rm Roche} = 146 R_{\odot}$.
We then determined the disk effective radius over the wavelength grid,
and our results for $R_d / R_s$ are plotted in Figure~6.

\begin{figure}
\begin{center}
{\includegraphics[angle=90,height=12cm]{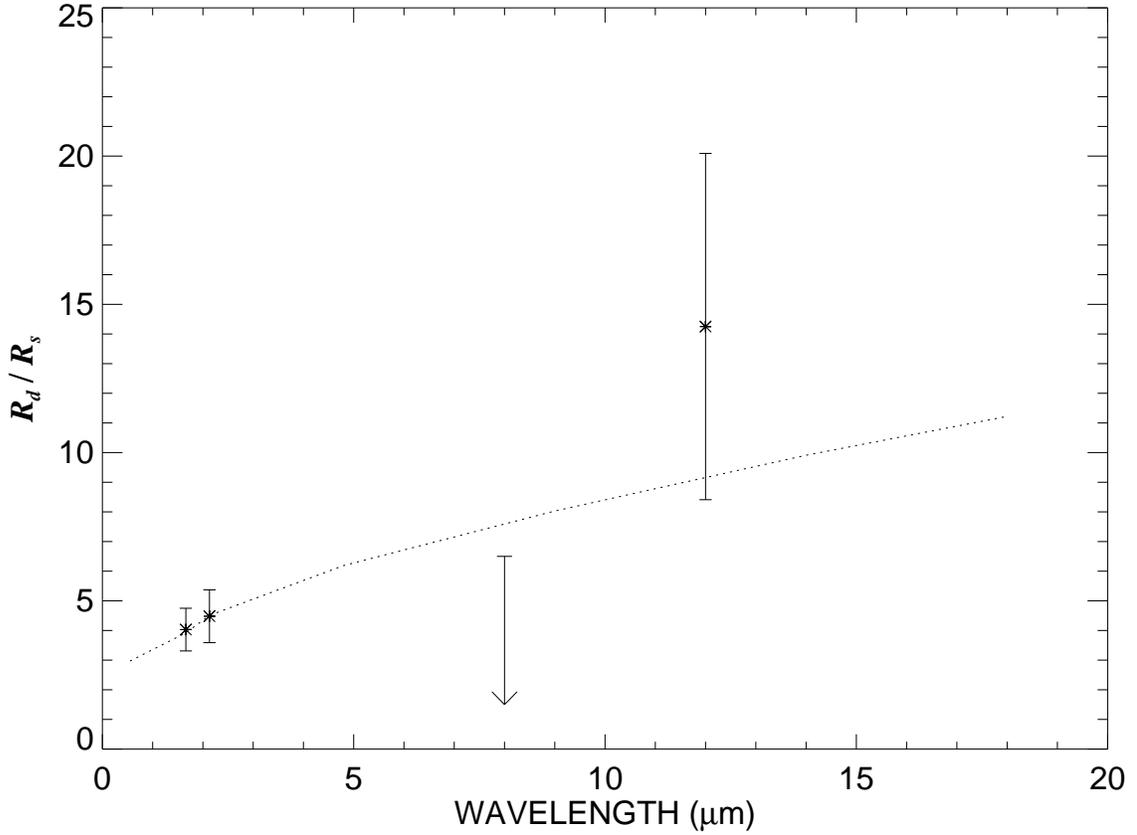}}
\end{center}
\caption{The disk-to-stellar radius ratio as a function of wavelength
for our model of the Be star $\zeta$~Tau (dotted line).  Also shown
are the measured sizes from interferometry (or upper limit on the
size for the 8~$\mu$m observation) based on an adopted stellar
angular diameter of 0.40~mas.}
\label{fig6}
\end{figure}

\placefigure{fig6}  

In order to test our model predictions, we collected recent interferometric
measurements of the angular size of $\zeta$~Tau.  We start with the two
CHARA Array results. The weighted average of the $H$-band, Gaussian FWHM
from \citet{sch10} is $1.61 \pm 0.05$~mas, and the corresponding
$K^\prime$-band value is $1.79 \pm 0.07$~mas \citep{gie07}.
The target was also observed at longer wavelengths with VLTI/MIDI
by \citet{mei09}.  They found an upper limit of FWHM
less than $2.6$~mas at 8~$\mu$m, but they resolved the disk at 12~$\mu$m
and found a FWHM $= 5.7 \pm 2.2$~mas.
These estimates are over-plotted for comparison on the theoretical
dotted-line shown in Figure~6 by assuming a stellar, limb darkened,
angular diameter of $\theta_{LD} = 0.40 \pm 0.04$~mas.
We find that the size of the emitting region does increase with increasing
wavelength in a manner mostly consistent with the observations.
The 8~$\mu$m upper limit falls slightly below the model curve in Figure~6,
but we suspect that this difference is marginal given the fact
that the VLTI/MIDI observations were not made at an optimal position
angle for the disk's orientation in the sky.  Thus, we suggest that
the viscous decretion disk predictions about disk size as a function of
wavelength pass the test for the case of $\zeta$~Tau.  However, it
is certainly possible that disk truncation effects will be more
important in binary Be stars with smaller semimajor axes.

We are currently completing a survey of bright, northern sky, Be stars
with the CHARA Array in the $K^\prime$-band, and we plan to use
our model to fit both the angular diameters and flux excesses.
Since the flux excess depends on the angular size and source function,
we will be able to infer the source function and hence
disk temperature for a diverse sample of Be stars.  We also intend
to compare the $H$ and $K^\prime$-band diameters for a subset
of Be stars with CHARA MIRC observations \citep{sch10}.


\acknowledgments

This material is based on work supported by the
National Science Foundation under Grant AST-0606861.
YT thanks the NASA Georgia Space Grant Consortium for a fellowship.
Institutional support has been provided from the GSU College
of Arts and Sciences and from the Research Program Enhancement
fund of the Board of Regents of the University System of Georgia,
administered through the GSU Office of the Vice President for Research.
We gratefully acknowledge all this support.

{\it Facility:} \facility{CHARA}



\clearpage

\begin{deluxetable}{lccccccc}
\tablewidth{0pc}
\tabletypesize{\scriptsize}
\tablenum{1}
\tablecaption{IR Color Excesses for Different Viscous Disk Models\label{tab1}}
\tablehead{
\colhead{} & \multispan{7}{\hfil $\rho_0$ (g~cm$^{-3}$) \hfil} \\
\colhead{Parameter} &
\colhead{$1.0\times 10^{-12}$} &
\colhead{$5.1\times 10^{-12}$} &
\colhead{$1.0\times 10^{-11}$} &
\colhead{$3.1\times 10^{-11}$} &
\colhead{$8.1\times 10^{-11}$} &
\colhead{$1.0\times 10^{-10}$} &
\colhead{$2.0\times 10^{-10}$}
}
\startdata
\multispan{8}{\hfil
$n = 3$, $i = 45^\circ$, $T_{\rm eff} = 30$ kK, $R_{\rm out} = 21.4 R_s$, $T_d = 2/3 ~T_{\rm eff}$
\hfil} \\
\hline
$E^*(V^*-H) $          \dotfill & 0.02 & 0.03 & 0.06 & 0.29 & 0.91 & 1.08 & 1.65 \\
$E^*(V^*-K)$           \dotfill & 0.02 & 0.05 & 0.11 & 0.59 & 1.38 & 1.57 & 2.19 \\
$E^*(V^*-4.8~\mu$m)    \dotfill & 0.02 & 0.11 & 0.31 & 1.04 & 1.86 & 2.06 & 2.69 \\
$E^*(V^*-9~ \mu$m)     \dotfill & 0.04 & 0.43 & 0.91 & 1.86 & 2.76 & 2.97 & 3.61 \\
$E^*(V^*-18~\mu$m)     \dotfill & 0.13 & 1.00 & 1.61 & 2.59 & 3.51 & 3.71 & 4.33 \\
$R_d /R_s$($H$)        \dotfill & 1.23 & 1.24 & 1.25 & 1.41 & 2.10 & 2.28 & 2.97 \\
$R_d /R_s$($K$)        \dotfill & 1.23 & 1.24 & 1.26 & 1.58 & 2.39 & 2.61 & 3.40 \\
$R_d /R_s$($4.8~\mu$m) \dotfill & 1.24 & 1.27 & 1.39 & 2.17 & 3.15 & 3.44 & 4.56 \\
$R_d /R_s$($9~\mu$m)   \dotfill & 1.24 & 1.38 & 1.79 & 2.76 & 4.07 & 4.49 & 6.05 \\
$R_d /R_s$($18~\mu$m)  \dotfill & 1.26 & 1.95 & 2.55 & 3.92 & 5.95 & 6.57 & 8.82 \\
\hline
\multispan{8}{\hfil
$n = 3$, $i = 80^\circ$, $T_{\rm eff} = 30$ kK, $R_{\rm out} = 21.4 R_s$, $T_d = 2/3 ~T_{\rm eff}$
\hfil} \\
\hline
$E^*(V^*-H) $          \dotfill & 0.02 & 0.03 & 0.04 & 0.11 & 0.39 & 0.49 & 0.91 \\
$E^*(V^*-K) $          \dotfill & 0.02 & 0.03 & 0.04 & 0.25 & 0.72 & 0.87 & 1.41 \\
$E^*(V^*-4.8~\mu$m)    \dotfill & 0.02 & 0.05 & 0.13 & 0.48 & 1.11 & 1.28 & 1.88 \\
$E^*(V^*-9~\mu$m)      \dotfill & 0.03 & 0.19 & 0.43 & 1.11 & 1.95 & 2.15 & 2.78 \\
$E^*(V^*-18~\mu$m)     \dotfill & 0.07 & 0.49 & 0.89 & 1.78 & 2.68 & 2.88 & 3.48 \\
$R_d /R_s$($H$)        \dotfill & 1.24 & 1.26 & 1.31 & 1.83 & 2.77 & 2.95 & 3.99 \\
$R_d /R_s$($K$)        \dotfill & 1.24 & 1.27 & 1.39 & 2.14 & 3.13 & 3.49 & 4.56 \\
$R_d /R_s$($4.8~\mu$m) \dotfill & 1.24 & 1.39 & 1.80 & 2.87 & 4.29 & 4.62 & 6.21 \\
$R_d /R_s$($9~\mu$m)   \dotfill & 1.26 & 1.77 & 2.43 & 3.64 & 5.61 & 6.11 & 8.17 \\
$R_d /R_s$($18~\mu$m)  \dotfill & 1.37 & 2.57 & 3.43 & 5.33 & 8.03 & 8.86 & 11.97\phn \\
\hline
\multispan{8}{\hfil
$n = 3$, $i = 45^\circ$, $T_{\rm eff} = 15$ kK,$R_{\rm out} =21.4 R_s$, $T_d = 2/3 ~T_{\rm eff}$
\hfil} \\
\hline
$E^*(V^*-H) $          \dotfill & 0.02 & 0.04 & 0.08 & 0.43 & 1.11 & 1.28 & 1.86 \\
$E^*(V^*-K) $          \dotfill & 0.02 & 0.06 & 0.15 & 0.73 & 1.53 & 1.73 & 2.35 \\
$E^*(V^*-4.8~\mu$m)    \dotfill & 0.04 & 0.17 & 0.46 & 1.26 & 2.11 & 2.31 & 2.94 \\
$E^*(V^*-9~\mu$m)      \dotfill & 0.05 & 0.49 & 1.01 & 1.96 & 2.86 & 3.07 & 3.71 \\
$E^*(V^*-18~\mu$m)     \dotfill & 0.14 & 1.04 & 1.64 & 2.63 & 3.54 & 3.79 & 4.36 \\
$R_d /R_s$($H$)        \dotfill & 1.23 & 1.24 & 1.27 & 1.64 & 2.46 & 2.68 & 3.51 \\
$R_d /R_s$($K$)        \dotfill & 1.23 & 1.25 & 1.29 & 1.77 & 2.62 & 2.85 & 3.76 \\
$R_d /R_s$($4.8~\mu$m) \dotfill & 1.24 & 1.30 & 1.55 & 2.44 & 3.59 & 3.94 & 5.23 \\
$R_d /R_s$($9~\mu$m)   \dotfill & 1.24 & 1.52 & 2.04 & 3.12 & 4.63 & 5.12 & 6.86 \\
$R_d /R_s$($18~\mu$m)  \dotfill & 1.27 & 1.99 & 2.62 & 4.03 & 6.07 & 6.71 & 9.03 \\
\hline
\multispan{8}{\hfil
$n = 3$, $i = 45^\circ$, $T_{\rm eff} = 30$ kK, $R_{\rm out} = 21.4 R_s$, $T_d = 1/2 ~T_{\rm eff}$
\hfil} \\
\hline
$E^*(V^*-H) $          \dotfill & 0.02 & 0.03 & 0.06 & 0.30 & 0.91 & 1.08 & 1.65 \\
$E^*(V^*-K)$           \dotfill & 0.02 & 0.04 & 0.09 & 0.45 & 1.16 & 1.34 & 1.94 \\
$E^*(V^*-4.8~\mu$m)    \dotfill & 0.03 & 0.11 & 0.31 & 1.04 & 1.86 & 2.06 & 2.69 \\
$E^*(V^*-9~\mu$m)      \dotfill & 0.04 & 0.31 & 0.72 & 1.61 & 2.48 & 2.68 & 3.33 \\
$E^*(V^*-18~\mu$m)     \dotfill & 0.09 & 0.79 & 1.35 & 2.32 & 3.22 & 3.42 & 4.05 \\
$R_d /R_s$($H$)        \dotfill & 1.23 & 1.24 & 1.25 & 1.41 & 2.10 & 2.28 & 2.97 \\
$R_d /R_s$($K$)        \dotfill & 1.23 & 1.24 & 1.26 & 1.55 & 2.34 & 2.54 & 3.31 \\
$R_d /R_s$($4.8~\mu$m) \dotfill & 1.24 & 1.27 & 1.39 & 2.17 & 3.15 & 3.44 & 4.56 \\
$R_d /R_s$($9~\mu$m)   \dotfill & 1.24 & 1.39 & 1.79 & 2.76 & 4.07 & 4.49 & 6.06 \\
$R_d /R_s$($18~\mu$m)  \dotfill & 1.26 & 1.86 & 2.46 & 3.75 & 5.67 & 6.25 & 8.48 \\
\hline
\multispan{8}{\hfil
$n = 3$, $i = 45^\circ$, $T_{\rm eff} = 30$ kK, $R_{\rm out} = 14.6 R_s$, $T_d = 2/3 ~T_{\rm eff}$
\hfil} \\
\hline
$E^*(V^*-H) $          \dotfill & 0.02 & 0.03 & 0.06 & 0.29 & 0.91 & 1.08 & 1.65 \\
$E^*(V^*-K) $          \dotfill & 0.02 & 0.05 & 0.12 & 0.60 & 1.38 & 1.57 & 2.18 \\
$E^*(V^*-4.8~\mu$m)    \dotfill & 0.02 & 0.11 & 0.31 & 1.04 & 1.86 & 2.06 & 2.67 \\
$E^*(V^*-9~\mu$m)      \dotfill & 0.05 & 0.43 & 0.92 & 1.86 & 2.75 & 2.96 & 3.57 \\
$E^*(V^*-18~\mu$m)     \dotfill & 0.13 & 1.00 & 1.59 & 2.58 & 3.47 & 3.66 & 4.23 \\
$R_d /R_s$($H$)        \dotfill & 1.23 & 1.24 & 1.25 & 1.42 & 2.10 & 2.29 & 2.97 \\
$R_d /R_s$($K$)        \dotfill & 1.23 & 1.24 & 1.27 & 1.58 & 2.39 & 2.60 & 3.40 \\
$R_d /R_s$($4.8~\mu$m) \dotfill & 1.24 & 1.27 & 1.39 & 2.17 & 3.15 & 3.44 & 4.56 \\
$R_d /R_s$($9~\mu$m)   \dotfill & 1.24 & 1.38 & 1.79 & 2.76 & 4.07 & 4.48 & 6.05 \\
$R_d /R_s$($18~\mu$m)  \dotfill & 1.26 & 1.95 & 2.55 & 3.92 & 5.95 & 6.56 & 8.82
\enddata
\end{deluxetable}


\begin{deluxetable}{cccc}
\tablewidth{0pt}
\tablenum{2}
\tablecaption{Adopted Main Sequence Colors\label{tab2}}
\tablehead{
\colhead{$T_{\rm eff}$}       &
\colhead{$V - K$}             &
\colhead{$V - 9~\mu{\rm m}$}  &
\colhead{$V - 18~\mu{\rm m}$} \\
\colhead{(kK)}   &
\colhead{(mag)}  &
\colhead{(mag)}  &
\colhead{(mag)}  }
\startdata
    10 & $-$0.06 & $-$0.07 & $-$0.07 \\
    12 & $-$0.23 & $-$0.29 & $-$0.29 \\
    14 & $-$0.34 & $-$0.43 & $-$0.44 \\
    16 & $-$0.43 & $-$0.54 & $-$0.55 \\
    18 & $-$0.51 & $-$0.64 & $-$0.65 \\
    20 & $-$0.58 & $-$0.74 & $-$0.75 \\
    22 & $-$0.64 & $-$0.82 & $-$0.83 \\
    24 & $-$0.70 & $-$0.90 & $-$0.91 \\
    26 & $-$0.74 & $-$0.96 & $-$0.98 \\
    28 & $-$0.79 & $-$1.02 & $-$1.03 \\
    30 & $-$0.83 & $-$1.07 & $-$1.08 \\
\enddata
\end{deluxetable}

\clearpage

\begin{deluxetable}{rccc}
\tabletypesize{\scriptsize}
\tablewidth{0pt}
\tablenum{3}
\tablecaption{Be Star Color Excesses\label{tab3}}
\tablehead{
\colhead{}                                   &
\colhead{$E^\star(V^\star-K) $}              &
\colhead{$E^\star(V^\star-9~\mu{\rm m}) $}   &
\colhead{$E^\star(V^\star-18~\mu{\rm m}) $}  \\
\colhead{HD}     &
\colhead{(mag)}  &
\colhead{(mag)}  &
\colhead{(mag)}  }
\startdata
    144 &    0.05     &    0.12     & \nodata \\
   4180 &    0.13     &    1.54     &    2.44 \\
   5394 &    0.54     &    1.83     &    2.51 \\
   6811 & $-$0.34\phs & $-$0.02\phs &    0.41 \\
  10144 &    0.03     &    0.29     &    0.43 \\
  10516 &    0.46     &    1.68     &    2.68 \\
  18552 &    0.11     &    0.74     &    1.62 \\
  20336 &    0.03     &    1.47     &    2.16 \\
  22192 &    0.33     &    1.75     &    2.62 \\
  23016 &    0.03     &    0.06     & \nodata \\
  23302 & $-$0.07\phs &    0.72     &    3.24 \\
  23480 & $-$0.06\phs &    0.29     &    1.72 \\
  23552 & $-$0.07\phs &    0.53     &    1.54 \\
  23630 &    0.29     &    0.28     &    0.97 \\
  23862 &    0.10     &    0.45     &    1.25 \\
  25940 &    0.08     &    1.16     &    1.99 \\
  28497 &    0.30     &    1.87     &    2.51 \\
  29866 &    0.03     &    0.72     &    1.49 \\
  30076 &    0.57     &    1.63     &    2.60 \\
  32343 &    0.39     &    1.21     &    2.43 \\
  32990 & $-$0.05\phs &    0.07     & \nodata \\
  32991 &    0.46     &    1.56     &    2.65 \\
  35439 & $-$0.23\phs &    1.54     &    2.12 \\
  36576 &    0.70     &    1.88     &    2.95 \\
  37202 &    0.72     &    1.71     &    2.56 \\
  37490 & $-$0.15\phs &    0.96     &    1.68 \\
  37795 &    0.10     &    0.51     &    1.30 \\
  41335 &    0.52     &    1.90     & \nodata \\
  44458 & $-$0.19\phs &    1.49     &    2.21 \\
  45542 &    0.00     &    0.32     &    0.89 \\
  45910 &    1.02     &    1.79     &    2.15 \\
  46860 &    0.06     &    0.13     & \nodata \\
  50013 &    0.23     &    1.49     &    2.21 \\
  50123 &    0.90     &    1.60     &    2.08 \\
  50820 &    1.01     &    0.80     &    1.33 \\
  54309 &    0.39     &    1.00     &    1.90 \\
  56014 & $-$0.04\phs &    0.88     &    1.40 \\
  56139 &    0.01     &    1.28     &    2.36 \\
  57150 &    0.40     &    1.67     &    2.52 \\
  57219 & $-$0.06\phs &    0.12     & \nodata \\
  58155 & $-$0.04\phs &    0.00     & \nodata \\
  58343 & $-$0.25\phs & \nodata     &    2.19 \\
  58715 & $-$0.03\phs &    0.47     &    1.17 \\
  60606 &    0.52     &    1.41     &    2.57 \\
  60855 &    0.75     & \nodata     &    3.09 \\
  63462 &    0.28     & \nodata     &    1.94 \\
  65875 &    0.39     &    1.73     &    2.64 \\
  66194 &    0.64     &    1.52     &    2.69 \\
  68980 &    0.48     &    1.62     &    2.45 \\
  71510 &    0.01     &    0.09     & \nodata \\
  72067 & $-$0.02\phs &    1.08     &    2.03 \\
  75311 & $-$0.09\phs &    0.19     &    1.05 \\
  77320 &    0.39     &    1.16     &    1.41 \\
  79621 &    0.03     &    0.02     & \nodata \\
  83953 &    0.50     & \nodata     &    2.37 \\
  86612 &    0.10     &    1.57     &    2.58 \\
  88661 &    0.54     &    1.78     &    2.69 \\
  91120 &    0.06     &    0.46     &    1.21 \\
  91465 &    0.32     &    1.51     &    2.23 \\
 105435 &    0.29     &    1.64     &    2.49 \\
 107348 &    0.03     &    0.55     & \nodata \\
 109387 &    0.29     &    1.42     &    2.30 \\
 110432 &    0.26     &    1.82     &    2.50 \\
 113120 & $-$0.44\phs &    1.33     &    2.09 \\
 120324 &    0.07     &    1.69     &    2.24 \\
 120991 &    0.29     &    0.07     &    1.03 \\
 121847 & $-$0.03\phs &    0.00     & \nodata \\
 124367 &    0.42     &    1.56     &    2.50 \\
 127972 &    0.10     &    1.13     &    1.87 \\
 137387 &    0.07     &    0.45     & \nodata \\
 138749 & $-$0.01\phs &    0.07     &    0.20 \\
 142184 &    0.09     &    0.54     &    1.18 \\
 142926 &    0.11     &    0.48     &    1.05 \\
 148184 &    0.49     &    1.93     &    2.71 \\
 153261 &    0.83     &    2.09     &    2.92 \\
 156325 &    0.00     &    0.12     & \nodata \\
 157042 &    0.20     &    1.34     &    2.02 \\
 158427 &    0.64     & \nodata     &    2.48 \\
 158643 &    0.39     & \nodata     &    4.40 \\
 164284 & $-$0.13\phs &    0.05     &    0.59 \\
 164447 & $-$0.07\phs &    0.12     & \nodata \\
 167128 &    0.03     &    0.62     &    1.34 \\
 168797 & $-$0.02\phs &    0.24     & \nodata \\
 170235 & $-$0.13\phs &    0.07     & \nodata \\
 171780 &    0.29     &    0.12     & \nodata \\
 173370 &    0.04     &    0.26     &    0.28 \\
 173948 &    0.12     & $-$0.03\phs &    0.23 \\
 174237 &    0.21     &    0.75     & \nodata \\
 175869 &    0.03     &    0.06     & \nodata \\
 178175 &    0.29     & \nodata     &    2.40 \\
 183362 &    0.55     &    1.77     & \nodata \\
 185037 &    0.04     &    0.61     & \nodata \\
 187567 &    0.56     &    1.78     &    2.63 \\
 187811 &    0.33     &    0.35     &    1.38 \\
 189687 &    0.04     &    0.33     &    1.46 \\
 191610 &    0.25     &    0.14     &    0.78 \\
 192044 &    0.12     &    0.81     &    1.37 \\
 193911 &    0.00     &    0.51     &    1.23 \\
 194244 &    0.04     &    0.24     & \nodata \\
 194335 &    0.19     &    1.34     &    2.41 \\
 195554 &    0.03     &    0.17     & \nodata \\
 196712 & $-$0.04\phs &    0.53     & \nodata \\
 197419 &    0.28     &    0.88     & \nodata \\
 198183 & $-$0.09\phs & $-$0.08\phs &    0.28 \\
 199218 &    0.08     &    0.84     & \nodata \\
 200120 &    0.45     &    1.69     &    2.13 \\
 200310 & $-$0.03\phs &    0.26     &    1.79 \\
 202904 &    0.04     &    1.62     &    2.50 \\
 203025 & $-$0.32\phs & $-$0.12\phs & \nodata \\
 203467 &    0.58     &    1.91     &    3.47 \\
 205551 & $-$0.14\phs &    0.06     & \nodata \\
 205637 &    0.14     &    0.69     &    1.37 \\
 208057 & $-$0.06\phs &    0.03     & \nodata \\
 208682 & $-$0.28\phs & $-$0.14\phs & \nodata \\
 209014 &    0.04     &    0.62     &    1.32 \\
 209409 &    0.22     &    1.10     &    1.94 \\
 209522 & $-$0.01\phs & $-$0.06\phs & \nodata \\
 210129 & $-$0.09\phs &    1.24     &    1.99 \\
 212076 &    0.41     &    1.74     &    2.48 \\
 212571 & $-$0.41\phs &    0.51     &    0.91 \\
 214168 &    0.25     &    0.62     & \nodata \\
 214748 & $-$0.02\phs &    0.14     &    0.99 \\
 216057 & $-$0.07\phs &    0.01     & \nodata \\
 216200 &    0.25     &    0.63     & \nodata \\
 217050 &    0.28     &    1.65     &    2.58 \\
 217543 & $-$0.07\phs &    0.33     & \nodata \\
 217675 & $-$0.12\phs &    0.04     &    0.44 \\
 217891 & $-$0.05\phs &    0.93     &    1.58 \\
 224544 &    0.00     & $-$0.07\phs & \nodata \\
 224559 &    0.10     &    1.52     &    2.15 \\
\enddata
\end{deluxetable}



\clearpage


\begin{thebibliography}{}
\bibitem[Bonanos et al.(2010)]{bon10}
         Bonanos, A. Z., et al. 2010, \aj, 140, 416
\bibitem[Carciofi \& Bjorkman(2006)]{cab06}
         Carciofi, A. C., \& Bjorkman, J. E.
         2006, \apj, 639, 1081
\bibitem[Carciofi et al.(2009)]{car09}
         Carciofi, A. C., Okazaki, A. T., le Bouquin, J.-B.,
         \v{S}tefl, S., Rivinius, T., Baade, D., Bjorkman, J. E., \&
         Hummel, C. A. 2009, \aap, 504, 915
\bibitem[Carciofi et al.(2006)]{car06}
         Carciofi, A. C., et al. 2006, \apj, 652, 1617
\bibitem[Clarke et al.(2005)]{cla05}
         Clarke, A. J., Oudmaijer, R. D., \& Lumsden, S. L. 2005, \mnras, 363, 1111
\bibitem[Code et al.(1976)]{cod76}
         Code, A. D., Bless, R. C., Davis, J., \& Brown, R. H. 1976, \apj, 203, 417
\bibitem[Cot{\'e} \& Waters(1987)]{cot87}
         Cot{\'e}, J., \& Waters, L. B. F. M. 1987, \aap,176, 93
\bibitem[Dougherty et al.(1994)]{dou94}
	 Dougherty, S. M., Waters, L. B. F. M., Burki, G., Cot\'{e}, J.,
         Cramer, N., van Kerkwijk, M. H., \& Taylor, A. R.
         1994, \aap, 290, 609
\bibitem[Elias et al.(1997)]{eli97}
         Elias, N. M., II, et al. 1997, \apj, 484, 394
\bibitem[Fitzpatrick(1999)]{fit99}
         Fitzpatrick, E. L. 1999, \pasp, 111, 63
\bibitem[Floquet et al.(1989)]{flo89}
         Floquet, M., Hubert, A. M., Maillard, J. P., Chauville, J., \& Chatzichristou, H.
         1989, \aap, 214, 295
\bibitem[Fr\'{e}mat et al.(2005)]{fre05}
         Fr\'{e}mat, Y., Zorec, J., Hubert, A.-M., \& Floquet, M. 2005, \aap, 440, 305
\bibitem[Gehrz et al.(1974)]{geh74}
         Gehrz, R. D., Hackwell, J. A., \& Jones, T. W., 1974, \apj, 191, 675
\bibitem[Gies et al.(2007)]{gie07}
         Gies, D. R., et al. 2007, \apj, 654, 527
\bibitem[Ginestet \& Carquillat(2002)]{gin02}
         Ginestet, N., \& Carquillat, J. M. 2002, \apjs, 143, 513
\bibitem[Halonen et al.(2008)]{hal08}
         Halonen, R. J., et al. 2008, \pasp, 120, 498
\bibitem[Hummel \& Vrancken(2000)]{hum00}
         Hummel, W., \& Vrancken, M. 2000, \aap, 359, 1075
\bibitem[Ishihara et al.(2010)]{ish10}
         Ishihara, D., et al. 2010, \aap, 514, A1
\bibitem[Ita et al.(2010)]{ita10}
         Ita, Y., et al. 2010, \aap, 514, A2
\bibitem[Jones et al.(2008)]{jon08}
         Jones, C. E., Sigut, T. A. A., \& Porter, J. M. 2008, MNRAS, 386, 1922
\bibitem[Kastner \& Mazzali(1989)]{kas89}
         Kastner, J. H, \& Mazzali, P. A., 1989, \aap, 210,295
\bibitem[Lamers \& Waters(1984)]{lam84}
         Lamers, H. J. G. L. M., \& Waters, L. B. F. M. 1984, \aap, 136, 37
\bibitem[Lee et al.(1991)]{lee91}
         Lee, U., Osaki, Y., \& Saio, H. 1991, \aap, 244, L5
\bibitem[Lejeune \& Schaerer(2001)]{lej01}
         Lejeune, T., \& Schaerer, D. 2001, \aap, 366, 538
\bibitem[Meilland et al.(2009)]{mei09}
         Meilland, A., Stee, P., Chesneau, O., \& Jones, C. 2009, \aap, 505, 687
\bibitem[Meilland et al.(2007)]{mei07}
         Meilland, A., et al. 2007, \aap, 464, 59
\bibitem[Okazaki(2001)]{oka01}
         Okazaki, A. T. 2001, PASJ, 53, 119
\bibitem[Okazaki et al.(2002)]{oka02}
         Okazaki, A. T., Bate, M. R., Ogilvie, G. I., \& Pringle, J. E.
         2002. \mnras, 337, 967
\bibitem[Porter(1999)]{por99}
         Porter, J. M. 1999, \aap, 348, 512
\bibitem[Porter(2003)]{prt03}
         Porter, J. M. 2003, Be Star Newsl., 36, 6
\bibitem[Porter \&  Rivinius(2003)]{por03}
         Porter, J. M., \&  Rivinius, Th. 2003, \pasp, 115, 1153
\bibitem[Quirrenbach et al.(1997)]{qui97}
         Quirrenbach, A., et al. 1997, \apj, 479, 477
\bibitem[Rinehart et al.(1999)Rinehart, Houck, \& Smith]{rin99}
         Rinehart, S. A., Houck, J. R., \& Smith, J. D. 1999, \aj, 118, 2974
\bibitem[Ru\v{z}djak et al.(2009)]{ruz09}
         Ru\v{z}djak, D., et al. 2009, \aap, 506, 1319
\bibitem[Schaefer et al.(2010)]{sch10}
         Schaefer, G. H., et al. 2010, \aj, 140, 1838
\bibitem[Sigut et al.(2009)]{sig09}
         Sigut, T. A., McGill, M. A., \& Jones, C. E. 2009, \apj, 699, 1973
\bibitem[Skrutskie et al.(2006)]{skr06}
         Skrutskie, M. F., et al. 2006, \aj, 131, 1163
\bibitem[Slettebak(1988)]{sle88}
         Slettebak, A. 1988, \pasp, 100, 770
\bibitem[Stee \& Bittar(2001)]{ste01}
         Stee, P., \& Bittar, J. 2001, \aap, 367, 532
\bibitem[Sterken et al.(1994)]{ste94}
         Sterken, C., Vogt, N., \& Mennickent, R. 1994, \aap, 291, 473
\bibitem[Struve(1931)]{str31}
         Struve, O. 1931, \apj, 73, 94
\bibitem[Tanab\'{e} et al.(2008)]{tan08}
         Tanab\'{e}, T., et al. 2008, PASJ, 60, S375
\bibitem[Touhami et al.(2010)]{tou10}
         Touhami, Y., et al.\ 2010, \pasp, 122, 379
\bibitem[Tycner et al.(2004)]{tyc04}
         Tycner, C., et al. 2004, \aj, 127, 1194
\bibitem[Tycner et al.(2006)]{tyc06}
         Tycner, C., et al. 2006, \aj, 131,2710
\bibitem[van Leeuwen(2007)]{van07}
         van Leeuwen, F. 2007, \aap, 474, 653
\bibitem[Waters(1986)]{wat86}
	 Waters, L. B. F. M. 1986, \aap, 162, 121
\bibitem[Waters et al.(1987)Waters, Cot\'{e}, \& Lamers]{wat87}
	 Waters, L. B. F. M., Cot\'{e}, J., \& Lamers, H. J. G. L. M.
         1987, \aap, 185, 206
\bibitem[Waters \& Lamers(1984)]{wat84}
	 Waters, L. B. F. M., \& Lamers, H. J. G. L. M. 1984, \aaps, 57, 327
\bibitem[Waters et al.(2000)]{wat00}
         Waters, L. B. F. M., et al. 2000, in The Be Phenomenon in Early-Type Stars,
         IAU Coll. 175 (ASP Conf. Proc. 214), ed. M. A. Smith, H, F. Henrichs,
         \& J. Fabregat (San Francisco: ASP), 145
\bibitem[Woolf et al.(1970)]{woo70}
         Woolf, N. J., Stein, W. A., \& Strittmatter, P. A. 1970, \aap, 9, 252
\bibitem[Zhang et al.(2005)]{zha05}
         Zhang, P., Chen P. S., \& Yang, H. T. 2005, NewA, 10,325
\end{thebibliography}
\end{document}